\documentclass[12pt]{iopart}
\usepackage{iopams}
\usepackage{iopams}
\usepackage{amsbsy}
\usepackage{xcolor}
\usepackage{soul}
\usepackage{graphicx}

\newcommand{\spvec}[1]{\ensuremath{\bi{#1}}}
\newcommand{\unitvec}[1]{\ensuremath{\hat{\bi{#1}}}}
\newcommand{\sptensor}[1]{\ensuremath{\boldsymbol{\mathbf{#1}}}}

\newcommand{\colvec}[1]{\ensuremath{\mathrm{#1}}}

\newcommand{\commentout}[1]{{}}
\newcommand{\half}{\hbox{$1\over2$}}

\newcommand{\rv}{\spvec{r}}
\newcommand{\Ev}{\spvec{E}}

\newcommand{\Hv}{\spvec{H}}
\newcommand{\Pv}{\spvec{P}}
\newcommand{\Mv}{\spvec{M}}

\newcommand{\kv}{\spvec{k}}

\newcommand{\beq}{\begin{equation}}
\newcommand{\eeq}{\end{equation}}
\newcommand{\rmM}{\mathrm{M}}
\newcommand{\rmE}{\mathrm{E}}

\newcommand{\rmO}{\mathrm{O}}
\newcommand{\rmr}{\mathrm{r}}
\newcommand{\rml}{\mathrm{l}}
\newcommand{\CASR}{\mathcal{C}^{(\mathrm{ASR})}}

\begin{document}
\title{Cooperative resonance linewidth narrowing in a planar metamaterial}
\author{Stewart D. Jenkins and Janne Ruostekoski}
\address{School of Mathematics and Centre for Photonic Metamaterials,
  University of Southampton, Southampton SO17 1BJ, United Kingdom}
\begin{abstract}
  We theoretically analyze the experimental observations of a spectral
  line collapse in a metamaterial array of asymmetric split ring
  resonators [Fedotov {\it et al.}, Phys.\ Rev.\ Lett.\ {\bf 104}, 223901
  (2010)].
  We show that the ensemble of closely-spaced resonators exhibits
  cooperative response, explaining the observed system-size dependent
  narrowing of the transmission  resonance linewidth.
  We further show that this cooperative narrowing depends sensitively
  on the lattice spacing and that significantly stronger narrowing could
  be achieved in media with suppressed ohmic losses.
\end{abstract}

\date{\today}
\maketitle

\section{Introduction}
\label{sec:intro}

Resonant multiple scattering plays an
important role in mesoscopic wave phenomena.
Such phenomena can be
  realized with electromagnetic (EM) fields.
In the strong scattering regime, interference of different scattering
paths between discrete scatterers can result in, e.g.,  light
localization \cite{WiersmaNat1997, vanTiggelen99}--an effect analogous
to the Anderson localization of electrons in solids.
Metamaterials comprise artificially structured media of plasmonic
resonators interacting with EM fields.
Due to several promising phenomena, such as the possibility for
diffraction-free lenses resulting from negative refractive index
\cite{SmithEtAlSCI2004}, there has been a rapidly increasing interest
in fabrication and theoretical modeling of such systems.
Additionally, the discrete nature of closely-spaced resonators in typical
metamaterial arrays raises the possibility to observe strong
collective radiative effects in these systems.

\begin{figure}[tbp]
  \centering
  \includegraphics[width=0.46\columnwidth]{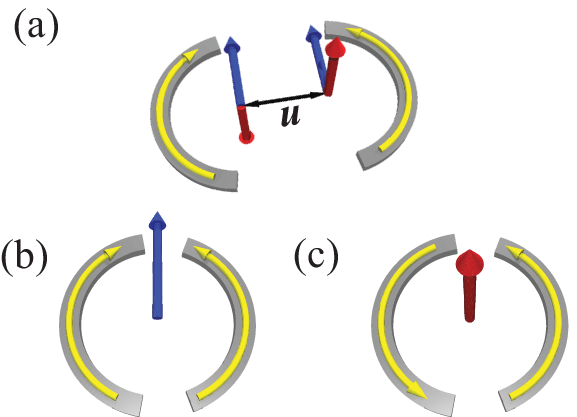}
  \includegraphics[width=0.46\columnwidth]{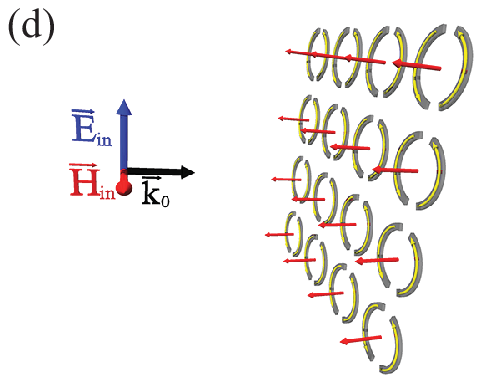}
  \caption{Asymmetric split ring (ASR) meta-molecules. (a) A
    schematic illustration of the two constituent meta-atoms of an ASR separated by distance
    $\spvec{u}$. Symmetric current oscillations
    produce parallel electric dipoles (blue arrows) but antiparallel
    magnetic dipoles (red arrows). (b) The
    symmetric mode with the currents in the two meta-atoms oscillating
    in-phase and (b) the antisymmetric mode with the current
    oscillating $\pi$ out-of-phase. For the symmetric case the
    dominant contribution is a net electric dipole moment in the plane
    of the resonator and for the antisymmetric case a net magnetic
    dipole moment normal to the plane of the resonator.
    (d) An
    illustration of an incident EM field driving the
    uniform phase-coherent collective magnetic eigenmode in which all the
    meta-molecules exhibit a magnetic dipole normal to the metamaterial
    plane. The incident field has an electric polarization along the
    ASR electric dipoles, but a magnetic field perpendicular to the
    ASR magnetic dipoles.
 } \label{fig1}
\end{figure}

In recent experiments Fedotov {\it et al.}\ observed a dramatic
suppression of radiation losses in a 2D planar metamaterial
array~\cite{FedotovEtAlPRL2010}.
The transmission spectra through the metamolecular sheet was found to
be strongly dependent upon the number of interacting meta-molecules in
the system.
The transmission resonance quality factor increased as a function of
the total number of active resonators, finally saturating at about 700
meta-molecules.
The metamaterial unit cell in the experiment was formed by an
asymmetric split-ring (ASR) resonator, consisting of two circular arcs
of slightly unequal lengths.
The currents in these ASRs may be excited symmetrically
(antisymmetrically), yielding a net oscillating electric (magnetic)
dipole as shown in figure~\ref{fig1}.

In this article we theoretically analyze the collective metamaterial
response, observed experimentally by Fedotov
  \textit{et al}~\cite{FedotovEtAlPRL2010}.
We find that strong interactions between a discrete set of resonators,
mediated by the EM field, characterize the response of the ensemble
and results in collective resonance linewidths and frequencies.
We show how the cooperative response of sufficiently closely-spaced resonators is
responsible for the observed narrowing of the transmission resonance
linewidth (increasing quality factor) with the number of
resonators~\cite{FedotovEtAlPRL2010}.
In particular, the system exhibits a collective mode with an almost
purely magnetic excitation, uniform phase profile, and strongly
suppressed radiative properties with each ASR possessessing a nearly
equal magnetic dipole moment.
We show in detail how this mode can be excited by an incident plane
wave propagating perpendicular to the array through an electric dipole
coupling to an ASR, even when the magnetic dipole moments are oriented
parallel to the propagation direction.
We calculate the resonance linewidth of the phase-coherent collective
magnetic mode that narrows as a function of the number of
ASRs, providing an excellent agreement with experimental
observations.
At the resonance, and with appropriately chosen parameters, nearly all
the excitation can be driven into this mode.
Due to its suppressed decay rate, the transmission spectrum displays a
narrow resonance.
The linewidth is sensitive to the spacing of the unit-cell
resonators, with the closely-spaced ASRs exhibiting cooperative
response, due to enhanced dipole-dipole interactions.
We find that the narrowing is limited by the ohmic losses
of the ASR resonators and that a dramatically stronger narrowing could be
achieved with a media exhibiting suppressed ohmic losses.

Our analysis demonstrates how essential features of the collective
effects of the experiment in \cite{FedotovEtAlPRL2010} can be
captured by a simple, computationally efficient model, developed in \cite{JenkinsLongPRB}, in which we treat each meta-atom as a discrete scatterer, exhibiting a single mode
of current oscillation and possessing appropriate electric and
magnetic dipole moments.
Interactions with the EM field then determine the collective
interactions within the ensemble.
Moreover, our analysis indicates the necessity of accounting for the
strong collective response of metamaterial systems and interference
effects in multiple scattering between the resonators in understanding
the dynamics and design of novel meta-materials.
Strong interactions between resonators can find important applications
in metamaterial systems, providing, e.g., precise control and
manipulation of EM fields on a sub-wavelength
scale~\cite{SentenacPRL2008,LemoultPRL10,KAO10}, in developments of a
lasing spaser~\cite{ZheludevEtAlNatPhot2008}, and disorder-related
phenomena~\cite{papasimakis2009,SavoEtAlPRB2012}.
Some features of interacting discrete resonators, such
  as the propagation of excitations through a 1D chain of
  meta-molecules \cite{LiuEtAlPRL2006}, could
  be modelled by introducing a phenomenological coupling between
  nearest neighbours into the Lagrangian describing metamaterial
  \cite{LiuEtAlPRL2006,LiuEtAlPRB2007}.
  Capturing the emergence of
  superradiant and subradiant collective mode linewidths, however,
  requires one to consider the repeated emission and
  reabsorption of radiation between resonators as described in
  \cite{JenkinsLongPRB}.
  Radiative coupling between pairs of magnetoelectric scatterers has
  also been considered in \cite{SersicEtAlPRB2011}.

The remainder of this article is organized as follows.
Section~\ref{sec:descr-reson-model} reviews the essential features of the theoretical
model developed in \cite{JenkinsLongPRB} that are required to analyze the
transmission resonance experiments~\cite{FedotovEtAlPRL2010}. We describe the metamaterial
as an ensemble of
discrete scatters, or meta-atoms, that dynamically respond to the EM
field.
In the context of this model, we describe the ASR meta-molecule that
forms the unit-cell of our metamaterial in
section~\ref{sec:asymm-split-ring}.
In section~\ref{sec:results:-linew-narr} we show how cooperative
interactions lead to the formation of the uniform magnetic mode in
which all ASR magnetic dipoles oscillate in phase.
We demonstrate how the
quality factor of this mode increases with the size of the system.
We illustrate how to excite this mode and compare the results of our
model to the experimental observations of Fedotov \textit{et al}.,
demonstrating a remarkable agreement.
Conclusions follow in section~\ref{sec:conclusion}.

\section{Theoretical Model}
\label{sec:descr-reson-model}

In order to analyze the experimental observations of the transmission
spectra~\cite{FedotovEtAlPRL2010}, we employ the theoretical model
developed in~\cite{JenkinsLongPRB}.
The model provides a computationally efficient approach for the
studies of strong collective EM field
mediated interactions between resonators in large metamaterial arrays.
In this section, we provide a brief outline of the basic
results of the general formalism of~\cite{JenkinsLongPRB} that are
needed to analyze the collective features of the EM response observed
in experiments \cite{FedotovEtAlPRL2010}. A more detailed description
of the model is provided in~\cite{JenkinsLongPRB}. In the following
section we apply the theory specifically to an array of ASR
metamolecules.

We consider an ensemble of $N$ metamaterial unit elements,
meta-molecules, each formed by $n$ discrete meta-atoms, with the
position of the meta-atom $j$ denoted by $\spvec{r}_j$
($j=1,\ldots,n\times N$).
An external beam with electric field
$\spvec{E}_{\mathrm{in}}(\spvec{r},t)$ and magnetic field
$\spvec{H}_{\mathrm{in}}(\spvec{r},t)$ with frequency
$\Omega_0$
drives the ensemble.
We assume the extent of each meta-atom is much less than the
wavelength $\lambda = 2\pi c/\Omega_0$ of the incident light so that
we may treat meta-atoms as radiating dipoles and ignore the
higher-order multipole-field interactions.
Within each meta-atom, EM fields drive the motion of charge carriers resulting
in oscillating charge and current distributions.
For simplicity, we assume that each meta-atom $j$ supports a single
eigenmode of current oscillation governed by a dynamic variable
$Q_j(t)$ with units of charge.
Then the associated electric and magnetic dipole moment for the
meta-atoms are
\begin{eqnarray}
  \label{eq:eDipDef}
  \spvec{d}_j = Q_j h_j \unitvec{d}_j \,\textrm{,} \\
  \spvec{m}_j = I_j A_j\unitvec{m}_j \,\textrm{,}
\end{eqnarray}
respectively, where $\unitvec{d}_j$ and $\unitvec{m}_j$ are unit vectors
denoting the dipole orientations and $I_j(t) = \rmd Q_j / \rmd t$ is the current.
Here $h_j$ and $A_j$ are the corresponding proportionality
coefficients (with the units of length and area) which depend on the
specific geometry of the resonators.
In the dipole approximation the polarization and magnetization are
given in terms of the density of electric and magnetic dipoles
\begin{eqnarray}
  \label{eq:totalPolMagDens}
  \Pv(\rv)=\sum_j \Pv_j(\rv) \,\textrm{,} \\
  \Mv(\rv)=\sum_j \Mv_j(\rv) \, \textrm{,}
\end{eqnarray}
where the polarization and the magnetization of the resonator $j$ are
\begin{eqnarray}
  \spvec{P}_j(\rv,t) \approx \spvec{d}_j \delta(\rv-\rv_j) \,\textrm{,} \\
  \spvec{M}_j(\rv,t) \approx \spvec{m}_j \delta(\rv-\rv_j) \, \textrm{,}
\end{eqnarray}
respectively.

The incident EM field drives the excitation of the current oscillations,
generating an oscillating electric and magnetic dipole in each
meta-atom.
The resulting dipole radiation from the metamaterial array is the sum
of the scattered electric and magnetic fields from all the meta-atoms
\begin{eqnarray}
  \label{eq:TotalScatteredFields}
  \Ev_{\mathrm{S}}(\rv,t)=\sum_j \Ev_{\mathrm{S},j}(\rv,t) \,\textrm{,} \\
  \Hv_{\mathrm{S}}(\rv,t)=\sum_j \Hv_{\mathrm{S},j}(\rv,t) \, \textrm{,}
\end{eqnarray}
where $\Ev_{\mathrm{S},j}(\rv,t)$ and $\Hv_{\mathrm{S},j}(\rv,t)$
denote the electric and magnetic field emitted by the meta-atom $j$.
The Fourier components of the scattered fields have the familiar
expressions of electric and magnetic fields radiated by oscillating
electric and magnetic dipoles~\cite{Jackson},
\begin{eqnarray}
  \fl \spvec{E}^+_{\mathrm{S},j}(\spvec{r},\Omega) =
  \frac{k^3}{4\pi\epsilon_0} \int \rmd^3 r' \,
  \Bigg[\sptensor{G}(\spvec{r} - \spvec{r}',\Omega) \cdot
    \spvec{P}^+_{j}(\spvec{r}',\Omega)    \nonumber\\
    + \frac{1}{c}
    \sptensor{G}_\times(\spvec{r}-\spvec{r}',\Omega) \cdot
    \spvec{M}^+_j(\spvec{r}',\Omega) \Bigg] , \label{efield}\\
    \fl \spvec{H}^+_{\mathrm{S},j}(\spvec{r},\Omega) =
  \frac{k^3}{4\pi} \int \rmd^3 r' \,
  \Big[\sptensor{G}(\spvec{r} - \spvec{r}',\Omega) \cdot
  \spvec{M}_j(\spvec{r}',\Omega) \nonumber\\
  - c \sptensor{G}_\times(\spvec{r}-\spvec{r}',\Omega) \cdot
  \spvec{P}^+_j(\spvec{r}',\Omega) \Big]\,,\label{mfield}
\end{eqnarray}
where we have defined the positive and negative frequency components
of a time varying real
quantity $V(t)$ such that for a Fourier component of
frequency $\Omega$ ($k\equiv \Omega/ c$), $V^{\pm}(\Omega) \equiv \Theta(\pm\Omega)
V(\Omega)$, and hence $V(t) = V^+(t) +V^{-}(t)$ with $V^{-}(t) =
[V^{+}(t)]^*$.
Here $\sptensor{G} (\spvec{r} - \spvec{r}',\Omega)$ denotes the
radiation kernel representing the electric (magnetic) field observed
at $\spvec{r}$ that is emitted from an electric (magnetic) dipole
residing at $\spvec{r}'$.
The radiation kernel that represents the magnetic (electric) field at $\spvec{r}$ scattered from an electric (magnetic) dipole source residing at $\spvec{r}'$ is denoted by
$\sptensor{G}_{\times} (\spvec{r} - \spvec{r}',\Omega)$.
Explicit expressions for $\sptensor{G}$ and $\sptensor{G}_\times$
coincide with the standard formulas of electromagnetism describing
dipole radiation~\cite{Jackson}.

Equations (\ref{efield}) and (\ref{mfield}) provide the total electric and magnetic fields
as a function of polarization and magnetization densities that are produced by current excitations in
the meta-atoms. In general, however, there is no simple way of solving
for $\spvec{P}(\spvec{r})$ and $\spvec{M}(\spvec{r})$. The scattered fields from each meta-atom drive the dynamics of the other meta-atoms in the system, with the EM fields mediating
interactions between the resonators. The radiated fields and the resonator excitations form a strongly coupled system when the separation between the resonators is of the order of the wavelength or less.

In order to solve the dynamics for the polarization and magnetization
densities appearing in equations (\ref{efield}) and (\ref{mfield}), we
have derived a coupled set of equations for the EM fields and
resonators~\cite{JenkinsLongPRB}. In the metamaterial sample, current
excitations in each meta-atom $j$ exhibit behaviour similar to that of
an LC circuit with resonance frequency
\begin{equation}
  \label{eq:resFreqDef}
  \omega_j \equiv \frac{1}{\sqrt{L_j C_j}} \,\textrm{,}
\end{equation}
where $C_j$ and $L_j$ denote effective self-capacitance and
self-inductance, respectively.
In this work we consider asymmetric meta-molecules consisting of two
meta-atoms with different resonance frequencies $\omega_j$, centered
around the frequency $\omega_0$, with $|\omega_j-\omega_0| \ll
\omega_0$.
The oscillating electric and magnetic dipoles radiate energy from an
isolated meta-atom at respective rates $\Gamma_{\rmE}$ and
$\Gamma_{\rmM}$ \cite{JenkinsLongPRB}
resulting in the scattered fields $\spvec{E}_{\mathrm{S},j}$ and
$\spvec{H}_{\mathrm{S},j}$. [See Equations~(\ref{efield}) and
(\ref{mfield}).]
The strengths of these radiative emission rates vary with the squares of the electric and magnetic dipole proportionality coefficients $h_j$ and  $A_j$, respectively \cite{JenkinsLongPRB}.
We assume that the meta-atom resonance
frequencies dominate the emission rates and that the resonance
frequencies occupy a narrow bandwidth around the dominant frequency of
the incident field, i.e., $\Gamma_{\rmE,j},\Gamma_{\rmM,j}, |\omega_0 -
\Omega_0| \ll \Omega_0$.
For simplicity, we also assume that the radiative electric and magnetic
decay rates of each resonator $\Gamma_{E}$ and $\Gamma_{M}$ are
independent of the resonator $j$.

The dynamics of current excitations in the meta-atom $j$ may then be
described by $Q_j(t)$ [introduced in equation~(\ref{eq:eDipDef})] and
its conjugate momentum $\phi_j(t)$ (with units of magnetic
flux)\cite{JenkinsLongPRB}.
In the absence of radiative emission and interactions with external fields,
the LC circuit, with resonance frequency $\omega_j$, formed by the oscillating charge and current can be
naturally described by
the slowly varying normal variables
\begin{equation}
  \label{eq:normalVariables}
  b_j(t) \equiv \frac{\rme^{\rmi \Omega_0 t}}{\sqrt{2}}
  \left(\frac{Q_j(t)}{\sqrt{\omega_jC_j}} +\rmi
    \frac{\phi_j(t)}{\sqrt{\omega_jL_j}}\right) \, \textrm{.}
\end{equation}
These normal variables are defined such that
$b_j$ undergoes a phase
modulation with frequency $(\omega_j - \Omega_0)$, i.e. $b_j(t) = b_j(0)
\exp\left[-i \left(\omega_j - \Omega_0\right) t\right]$
which is perturbed by nonzero radiative losses $\Gamma_{\mathrm{E}},\Gamma_{\mathrm{M}} \ll \Omega_0$,
 and driving from the external fields.

The fields generated externally to each meta-atom $j$,
composed of the incident field and fields scattered from all
other meta-atoms in the system,
drive the amplitude $b_j$
of current oscillation within the meta-atom.
The component of the external electric field
  oriented along the dipole direction $\unitvec{d}_j$ provides a net
  external electromotive force (EMF),  and the component of the
  external magnetic field along the magnetic dipole direction
  $\unitvec{m}_j$ provides a net applied magnetic flux.
  The applied external EMF and oscillating magnetic flux
  induce current flow in the meta-atom.
The oscillating current of that meta-atom, in turn, generates electric
and magnetic dipoles which both radiate magnetic and electric fields,
according to (\ref{efield}) and~(\ref{mfield}).
These fields couple to dynamical variables of charge oscillations in
other meta-atoms, producing more dipolar radiation.
The result of this multiple scattering is that the EM fields mediate
interactions between the meta-atom dynamic variables.

For the limits we consider in this article, the
metamaterial's response to the incident EM field is then governed by
the set of coupled linear equations for the meta-atom variables $b_j$
\cite{JenkinsLongPRB},
\begin{equation}
  \dot{\colvec{b}} = \mathcal{C} \colvec{b} + \colvec{f}_{\mathrm{in}} \textrm{ ,}
  \label{eq:rwa_b_eqm}
\end{equation}
where we have introduced the notation for column vectors of normal
variables $\colvec{b}$ and the driving $\colvec{f}_{\mathrm{in}}$
caused by the incident field
\begin{equation}
  \label{eq:bColvec}
  \colvec{b}(t) \equiv \left(
    \begin{array}{c}
      b_1(t)\\
      b_2(t)\\
      \vdots\\
      b_{nN}(t)
    \end{array}
  \right) \, \textrm{,} \qquad
  \colvec{f}_{\mathrm{in}}(t) \equiv \left(
    \begin{array}{c}
      f_{1,\mathrm{in}} (t) \\
      f_{2,\mathrm{in}}(t) \\
      \vdots \\
      f_{nN,\mathrm{in}}(t)
    \end{array}
  \right) \,\textrm{.}
\end{equation}
The applied incident fields induce an EMF and magnetic flux in each
meta-atom $j$, producing the driving $f_{j,\mathrm{in}}$
\cite{JenkinsLongPRB}.
Under the experimental conditions we consider here, however,
the meta-atom magnetic dipoles are aligned perpendicular to the incident
magnetic field, and thus only the EMF contributes to the driving of
each meta-atom, which is given by
\begin{equation}
  \label{eq:f_driving}
  \rme^{-\rmi \Omega_0 t} f_{j,\mathrm{in}}(t) =
  \rmi
  \frac{h_j}{\sqrt{2\omega_j L_j}}
  \unitvec{d}_j \cdot
  \spvec{E}_{\mathrm{in}}^+(\rv_j,t)
 \, \textrm{.}
\end{equation}
The current oscillations excited by the incident electric field then
simultaneously produce electric and magnetic dipoles which scatter
fields to other meta-atoms, which then rescatter the fields.
This multiple scattering between the meta-atoms results in the linear
coupling matrix
\begin{equation}
  \mathcal{C}  =  -\rmi \mathrm{\Delta} - \frac{\Gamma}{2} \mathrm{I}
  + \frac{1}{2} \left[ \rmi
    \Gamma_{\mathrm{E}}
    \mathcal{G}_{\mathrm{E}} +
    \rmi
    \Gamma_{\mathrm{M}}
    \mathcal{G}_{\mathrm{M}}
    +
    \bar{\Gamma} \left(
    \mathcal{G}_\times
    +
   \mathcal{G}_\times^{T} \right) \right] \textrm{,}
  \label{eq:C_rwa}
\end{equation}
where $\mathrm{I}$ represents the identity matrix, and $\bar{\Gamma}
\equiv \sqrt{\Gamma_{\rmE} \Gamma_{\rmM}}$ is the geometric mean of
the electric and magnetic dipole emission rates.
Here the detunings of the incident field from the meta-atom resonances
are contained in the diagonal matrix $\mathrm{\Delta}$ with elements
\begin{equation}
  \Delta_{j,j'} \equiv \delta_{j,j'} \left(\omega_j-\Omega_0\right)\,\textrm{,}
  \label{eq:DeltaMatDef}
\end{equation}
and the energy carried away from individual meta-atoms by the
scattered fields manifests itself in the decay rate \cite{JenkinsLongPRB}
\begin{equation}
  \label{eq:GammaMatDef}
  \Gamma \equiv \Gamma_{\rmE} +
    \Gamma_{\rmM} + \Gamma_{\mathrm{O}}
\end{equation}
appearing in the diagonal elements of $\mathcal{C}$.
In the limits we consider here, a meta-atom's magnetic and
electric dipoles oscillate $\pi/2$ out of
phase~\cite{JenkinsLongPRB} with one and other.
The fields radiated from a single meta-atom's electric and magnetic
dipoles therefore neither constructively nor destructively interfere
with each other, and an  isolated meta-atom's radiative emission rate
is the sum of $\Gamma_{\mathrm{E}}$ and
$\Gamma_{\mathrm{M}}$~\cite{JenkinsLongPRB}.
In addition to the radiative losses, we have included a
phenomenological rate $\Gamma_{\mathrm{O}}$ to account for non-radiative,
e.g. ohmic losses.
The inter-meta-atom interactions produced by the scattered fields
result in electric and magnetic dipole-dipole interactions, accounted
for by matrices $\mathcal{G}_{\rmE}$ and $\mathcal{G}_{\rmM}$,
respectively, that depend on the relative positions and orientations of the meta-atom dipoles (the precise form is given in \cite{JenkinsLongPRB}). 
Additionally, the \emph{electric} field emitted by the
\emph{magnetic} dipole of one meta-atom will drive the
electric dipole of another.
The resulting interaction reveals itself in the cross coupling matrix
$\mathcal{G}_\times$.
Similarly, the magnetic field emitted by the electric dipole of one
meta-atom interacts with the magnetic dipoles of others.
These interactions manifest themselves as $\mathcal{G}_\times^T$, the
transpose of the cross coupling matrix.
The interaction processes between the different resonators, mediated
by dipole radiation, are analogous to frequency dependent mutual
inductance and capacitance, but due to the radiative long-range
interactions, these can substantially differ from the quasi-static
expressions for which ${\mathcal{C}_\times}$ is also absent.

In order to calculate the EM response of the system, we solve the
coupled set of equations (\ref{eq:rwa_b_eqm}) involving all the
resonators and the fields.
A system of $n\times N$ single-mode resonators then possesses $n\times
N$ collective modes of current oscillation.
Each collective mode exhibits a distinct collective linewidth (decay
rate) and resonance frequency, determined by the imaginary and real
parts of the corresponding eigenvalue \cite{JenkinsLongPRB}.
The resulting dynamics resemble a cooperative response of atomic gases
to resonant light in which case the EM coupling between different
atoms is due to electric dipole radiation alone
\cite{MoriceEtAlPRA1995,RuostekoskiJavanainenPRA1997L,RuostekoskiJavanainenPRA1997,fermiline,JavanainenEtAlPRA1999,dalibardexp}.
The crucial component of the strong cooperative response of
closely-spaced scatterers are \emph{recurrent} scattering events
\cite{vantiggelen90,Ishimaru1978,RuostekoskiJavanainenPRA1997L,RuostekoskiJavanainenPRA1997,fermiline,JavanainenEtAlPRA1999}
-- in which a wave is scattered more than once by the same dipole.
Such processes cannot generally be modeled by the continuous medium
electrodynamics, necessitating the meta-atoms to be treated as
discrete scatterers.
An approximate calculation of local field corrections in a
magnetodielectric medium of discrete scatterers was performed in
\cite{Kastel07} where the translational symmetry of an infinite
lattice simplifies the response.

\section{The asymmetric split ring meta-molecule}
\label{sec:asymm-split-ring}

In this article, we provide a theoretical analysis of the experimental findings
by Fedotov \textit{et al} \cite{FedotovEtAlPRL2010} and explain the observed
linewidth narrowing of the transmission spectrum for a 2D metamaterial array of ASRs.
We will show that the observed transmission resonance
\cite{FedotovEtAlPRL2007} and its enhanced quality factor as a function of the size of the system result from
the formation of a collective mode whose decay rate becomes
more suppressed for increased array sizes.
Within our model,
a single ASR, consisting of $2$ meta-atoms, has $2$ modes of
oscillation, each of which decay at a rate comparable to the single,
isolated
meta-atom decay rate $\Gamma$.
It is only when the ASRs act in concert that the transmission
resonance due to linewidth narrowing can be observed.
In order to understand the collective dynamics of the metamaterial, we
first show how the EM mediated interactions between meta-atoms
determine the behaviour of its constituent meta-molecule.
To that end, it is instructive to first apply our theoretical model
to describe the behaviour of a single, isolated,
ASR.

A single ASR consists of two separate concentric circular arcs
(meta-atoms), labeled by $j\in\{\rml,\rmr\}$ (for ``left'' and ``right'')
as illustrated in figure~\ref{fig1}.
The ASR is an example of a split ring resonator, variations of which
are instrumental in the production of metamaterials with exotic
properties such as negative indices of refraction \cite{ShelbySci2001,
  SmithEtAlPRL2000}.
To illustrate the qualitative physical behaviour of the ASR, we
approximate the meta-atoms as two point sources located at points
$\rv_{\rmr}$ and $\rv_{\rml}$ separated by $\spvec{u} \equiv
\spvec{r}_{\rmr}-\spvec{r}_{\rml}$.
The current oscillations in each meta-atom produce electric dipoles
with orientation $\unitvec{d}_{\rmr} = \unitvec{d}_{\rml} \equiv
\unitvec{d}$, where $\unitvec{d}\perp\unitvec{u}$.
Owing to the curvature of each meta-atom, current oscillations produce
magnetic dipoles with opposite orientations $\unitvec{m}_{\rmr} =
-\unitvec{m}_{\rml} \equiv \unitvec{m}$, where $\unitvec{m}\perp\spvec{u}$ and $\unitvec{m}\perp\unitvec{d}$.
An asymmetry between the rings, in this case resulting from a difference in
arc length, manifests itself as a difference in resonance frequencies
with $\omega_{\rmr} = \omega_0 + \delta\omega$ and $\omega_{\rml} =
\omega_0 - \delta\omega$.

Although this simplified model does not account for higher order
multipole contributions of an individual meta-atom, oscillations  of
the ASR meta-molecule consisting of two meta-atoms does exhibit
non-vanishing quadrupole moments.
While this quadrupole contribution can be inaccurately represented in
the dipole approximation, in the case of an ASR modes, the electric
quadrupole moment is notably suppressed in most experimental
situations when compared to the corresponding dipolar field
\cite{PapasimakisComm}.
The dipole approximation also provides an advantage in computational
efficiency and in maintaining the tractability of the calculation.
Despite the dipole approximation implemented in the numerics we find
in section~\ref{sec:results:-linew-narr} that the model is able to
reproduce the experimental findings of the enhanced quality factor of
the transmission resonance observed by Fedotov \textit{et al}
\cite{FedotovEtAlPRL2010}.

The dynamics of a single, isolated, ASR are described by the two
normal variables $b_{\rmr}$ and $b_{\rml}$ for the right and left meta-atoms,
respectively.
These two meta-atoms interact via electric and magnetic
dipole-dipole interactions as well as interactions due to the electric
(magnetic) fields emitted by the other meta-atom's magnetic (electric)
dipole.
The dynamics of the ASR are given by [see equation (\ref{eq:rwa_b_eqm})]
\begin{equation}
  \label{eq:ASRDynamics}
  \frac{\rmd}{\rmd t} \left(
    \begin{array}{c}
      b_{\rmr}(t) \\
      b_{\rml}(t)
    \end{array}
  \right) =
  \CASR \left(
    \begin{array}{c}
      b_{\rmr}(t) \\
      b_{\rml}(t)
    \end{array}
  \right)
  +
  \left(
    \begin{array}{c}
      f_{\rmr,\mathrm{in}}(t) \\
      f_{\rml,\mathrm{in}}(t)
    \end{array}
  \right) \,\textrm{.}
\end{equation}
The driving terms $f_{\rmr,\mathrm{in}}$ and $f_{\rml,\mathrm{in}}$
are given in (\ref{eq:f_driving}), while the interaction matrix
\begin{equation}
  \label{eq:C_ASR}
  \CASR = \left(
    \begin{array}{cc}
      -\rmi \left(\omega_0 +\delta\omega - \Omega_0\right) -
      \Gamma/2 & i\left(\Gamma_{\mathrm{E}} -
        \Gamma_{\mathrm{M}}\right)G - \bar{\Gamma}S \\
     i\left(\Gamma_{\mathrm{E}} -
        \Gamma_{\mathrm{M}}\right)G - \bar{\Gamma}S & -\rmi \left(\omega_0
        -\delta\omega - \Omega_0\right) -
      \Gamma/2
    \end{array}
  \right) \,\textrm{.}
\end{equation}
The off-diagonal matrix elements are identical and account for the
EM interactions between the two
meta-atoms.
Because the meta-atoms oscillating in phase produce parallel
electric dipoles, but antiparallel magnetic dipoles, the magnetic
dipole-dipole interaction (proportional to $\Gamma_{\mathrm{M}}$)
differs in sign from the electric dipole dipole interactions
(proportional to $\Gamma_{\mathrm{E}}$) and have a strength related
to the meta-atom separation by the factor
\begin{equation}
  \label{eq:GUnboldDef}
  G \equiv \frac{3}{4} \unitvec{d}\cdot
  \sptensor{G}(\spvec{u},\Omega_0) \cdot \unitvec{d}
  = \frac{3}{4} \unitvec{m}\cdot
  \sptensor{G}(\spvec{u},\Omega_0) \cdot \unitvec{m} \,\textrm{.}
\end{equation}
The interaction between the electric (magnetic) dipole of one
meta-atom and the magnetic (electric) dipole of the other is
proportional to $\bar{\Gamma}$, and is associated with the
geometrical factor
\begin{equation}
  \label{eq:2}
  S \equiv \frac{3}{4} \unitvec{d}\cdot
  \sptensor{G}_\times(\spvec{u},\Omega_0) \cdot \unitvec{m} \,\textrm{.}
\end{equation}

\begin{figure}
  \centering
  \includegraphics{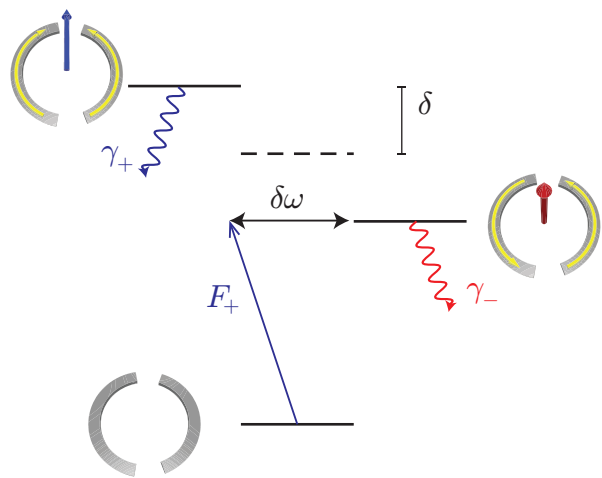}
  \caption{A diagram illustrating the symmetric and antisymmetric
    modes of an ASR driven by an electric field resonant on the
    antisymmetric mode.
    The resonance frequency of the symmetric (antisymmetric) mode is
    shifted up (down) from the central frequency $\omega_0$ by
    $\delta$.
    Radiative decay is illustrated by the decay rates $\gamma_\pm$ of
    the mode variables $c_\pm$.
    The asymmetry $\delta\omega$ manifests itself as a coupling
    between the symmetric and antisymmetric modes.
  }
  \label{fig:ASRLevels}
\end{figure}

In an isolated ASR the radiative interactions between the two
resonators result in eigenstates analogous to superradiant and
subradiant states in a pair of atoms.
In order to analyze these eigenstates, we consider the dynamics of
symmetric $c_+$ and antisymmetric $c_-$ modes of current oscillation
(figure~\ref{fig:ASRLevels}) that represent the exact eigenmodes of the ASR in
the absence of asymmetry $\delta\omega=0$ \cite{JenkinsLongPRB}.
In terms of the individual meta-atom normal variables the symmetric
and antisymmetric modes are given by
\begin{equation}
  \label{eq:c_pm}
  c_\pm(t) \equiv \frac{1}{\sqrt{2}} \left(b_{\rmr}(t) \pm
    b_{\rml}(t)\right) \,\textrm{.}
\end{equation}
Excitations of these modes possess respective net electric and
magnetic dipoles and will thus be referred to electric and magnetic
dipole excitations.
The split ring asymmetry $\delta\omega\ne0$, however, introduces an
effective coupling between these modes in a single ASR, so that
\begin{equation}
  \frac{\rmd{c}_\pm}{\rmd t} =
  \left[-\gamma_\pm /2 -\rmi\left(\omega_0 \pm \delta
      -\Omega_0\right)\right]c_\pm - \rmi\delta\omega c_\mp + F_\pm
  \, \textrm{,}
  \label{eq:c_pm_evolve}
\end{equation}
where $\gamma_\pm$ and $\delta$ denote the decay rates and a frequency
shift, respectively, and $F_\pm$ represents the driving by the
incident field.
The decay rates and frequency shifts of the ASR modes, which arise from the
inter-meta-atom interactions, are independent of the meta-atom
asymmetry, and their exact form is given in \cite{JenkinsLongPRB}.

When the spacing between meta-atoms is much less than a wavelength,
the symmetric mode decays entirely due to electric dipole radiation,
while the antisymmetric mode suffers decay from magnetic dipole
radiation resulting in the ASR mode decay rates
\begin{eqnarray}
  \label{eq:gammaPlus}
  \gamma_+ \approx 2\Gamma_{\rmE} + \Gamma_{\rmO} \, \textrm{,} \\
  \gamma_- \approx 2\Gamma_{\rmM} + \Gamma_{\rmO} \, \textrm{.}
  \label{eq:gammaMinus}
\end{eqnarray}
Furthermore, the symmetric and antisymmetric modes are driven purely
by the external electric and magnetic fields respectively, with $F_+
\propto \unitvec{d} \cdot \spvec{E}_{\mathrm{in}}(\spvec{R},t)$ and
$F_- \propto \unitvec{m} \cdot \spvec{H}_{\mathrm{in}}(\spvec{R},t)$
where $\spvec{R}$ is the centre of mass of the ASR.

The asymmetry $\delta\omega$ results in a coupling between the
symmetric and antisymmetric modes.
Figure~\ref{fig:ASRLevels} illustrates these modes as being analogous
to molecular excitations \cite{ProdanEtAlSCI2003,WangEtAlACR2006} with
an unexcited ASR represented by the ground state.
Consider an incident EM field whose magnetic field is perpendicular to
$\unitvec{m}$ so that it only drives the meta-molecule electric
dipoles.
In the absence of asymmetry, this incident field could only
drive the symmetric mode.
The coupling induced by the asymmetry, on the other hand, can allow
such an incident field to additionally excite the antisymmetric mode.
For example, figure \ref{fig:ASRLevels} illustrates an incident field
resonant on the antisymmetric mode (with $\Omega_0 =
\omega_0-\delta$), but which exclusively drives the symmetric mode.
The asymmetry, permits a resonant excitation of the antisymmetric
mode in a process analogous to a two photon atomic transition.

In the calculations of the properties of a single, isolated ASR
the precise form of the geometrical factors $G$ and $S$ are a
  result of treating the meta-atoms as point emitters.
  More exact expressions for these factors, which influence the
  coupling between meta-atoms, could be obtained by accounting for the
  finite spatial extent of the circuit elements
  \cite{JenkinsLongPRB}. Approximating each meta-atom as a point emitter has its greatest
  effect on the frequency shift $\delta$ between the symmetric and antisymmetric
  oscillation of individual meta-molecules.
The frequency shift $\delta$ depends on dipole-dipole interaction energies and is
  therefore very sensitive ($\sim 1/u^3$) to the inter-meta-atom spacing. On the other hand,
  the decay rates $\gamma_\pm$ are
  insensitive to the meta-atom spacing and would be unaltered if one
  did not approximate the meta-atoms as point emitters.
  As only the numerical coefficients would be affected,
  however, the physical behaviour described by
  equation~(\ref{eq:c_pm_evolve}) is unchanged in the dipole approximation.

\section{Theoretical analysis of the experimentally observed
  transmission resonance
linewidth narrowing}
\label{sec:results:-linew-narr}

In practice, one does not observe excitation of a single ASR's
magnetic dipole by an incident
field which solely drives the electric dipole.
This is because the radiative decay rate of a magnetic dipole excitation
in an isolated ASR is approximately as fast as the decay rate of an electric dipole excitation.
Any energy in the antisymmetric mode is therefore radiated away
before it can be appreciably excited.
We find that this changes dramatically when
many ASRs respond cooperatively.

An incident field driving only electric dipoles can excite a high
quality magnetic mode in which all ASR magnetic dipoles oscillate in
phase.
Excitation of this magnetic mode at the expense of other modes is
responsible for the transmission resonance observed by
Fedotov \textit{et al} \cite{FedotovEtAlPRL2007,FedotovEtAlPRL2010}.
We find that the collective decay rate of this mode decreases (the
quality factor increases) with an increasing
number of ASRs participating in the metamaterial.
Allowing for ohmic losses, our model provides excellent agreement with
the enhanced transmission resonance quality factor for increased
system size observed by Fedotov \textit{et al}
\cite{FedotovEtAlPRL2010}.

To model the experimentally observed collective
response \cite{FedotovEtAlPRL2010} we study an ensemble of identical
ASRs (with $\spvec{u}=u \unitvec{e}_x$ and $\spvec{d}=d
\unitvec{e}_y$; figure~\ref{fig1}) arranged in a 2D square lattice
within a circle of radius $r_c$, with lattice spacing $a$, and lattice
vectors $(a\unitvec{e}_x,a \unitvec{e}_y)$.
The sample is illuminated by a cw plane wave
\begin{equation}
\Ev_{\rm in}^+(\rv)=
\half {\cal E} \unitvec{e}_y e^{i\kv \cdot \rv},
\end{equation}
with $\spvec{k} =k_0
\unitvec{e}_z$, coupling to the electric dipole moments of the ASRs.
In the experimentally measured transmission resonance through such a
sheet~\cite{FedotovEtAlPRL2010} the number of active ASRs, $N$, was
controlled by decoupling the ASRs with $r > r_c$ from the rest of the
system with approximately circular shaped metal masks with varying
radii $r_c$.
The resonance quality factor increased with the total number of active ASRs,
saturating at about $N=700$.

The electric and magnetic fields scattered from each ASR impinge on
other ASRs in the metamaterial which, in turn, rescatter the fields.
Multiple scattering processes result in an interaction between all
the meta-atoms in the array, manifesting themselves in the dynamic
coupling matrix $\mathcal{C}$ (\ref{eq:C_rwa})
between the normal variables.
Collective modes of the metamaterial are represented by eigenvectors
of $\mathcal{C}$, with the $i$th eigenvector denoted by
$\colvec{v}_i$.
The corresponding eigenvalue $\lambda_i$ gives the collective mode
decay rate and resonance frequency shift
\begin{eqnarray}
  \label{eq:collectiveDecay}
  \gamma_i = - 2\mathrm{Re}\,\lambda_i,
  \\
  \label{eq:collectiveShift}
  \delta_i = -\mathrm{Im}\,\lambda_i - \left(\omega_0 -
    \Omega_0\right),
\end{eqnarray}
respectively.

\begin{figure}
  \centering
  \includegraphics[width=0.46\columnwidth]{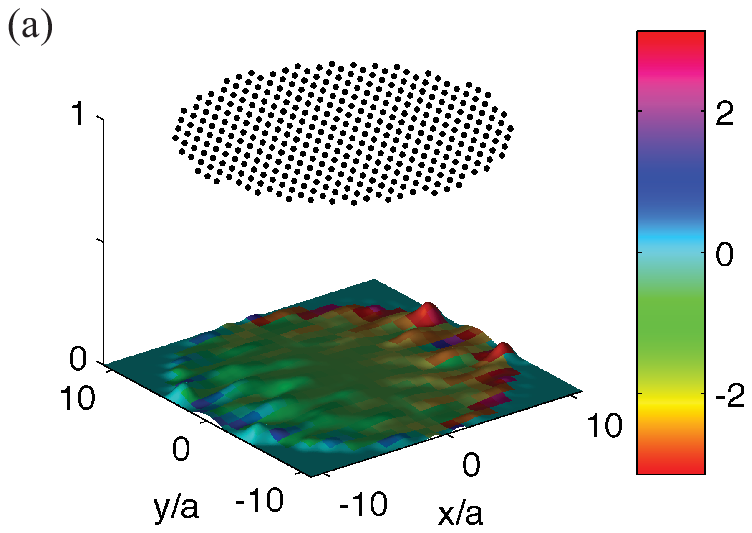}
  \includegraphics[width=0.46\columnwidth]{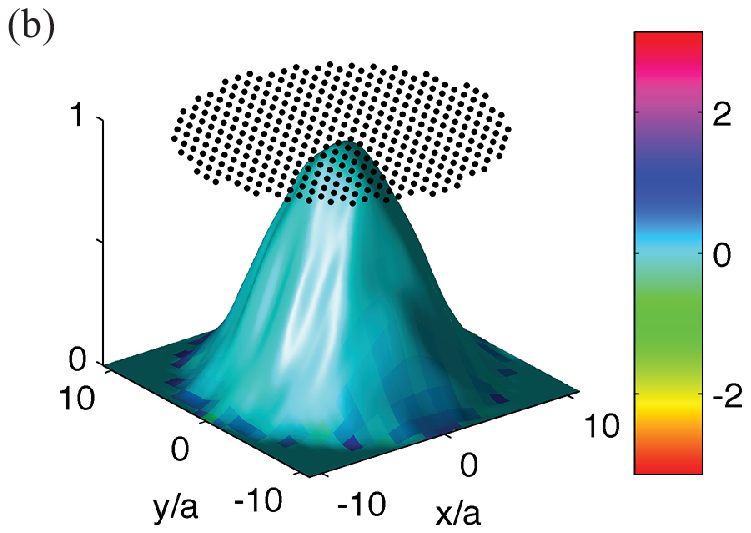}
  \caption{The numerically calculated uniform magnetic mode
    for an ensemble of 335 ASRs in which all magnetic dipoles
    oscillate in phase with minimal contribution from ASR electric
    dipole excitations.
    (a) The electric dipole excitation $|c_+|^2$
    and (b) the magnetic dipole
    excitations $|c_-|^2$ of the ASRs in the uniform magnetic mode
    $\colvec{v}_{\mathrm{m}}$.
    The phase of the electric ($c_+$) and magnetic ($c_-$) dipole
    excitations are indicated by the colour of the surfaces in (a)
    and (b) respectively.
    The black dots indicate the positions of the ASRs in the
    array.
    This mode was calculated for a lattice spacing of $a \simeq
    0.28 \lambda$ and an ASR asymmetry of $\delta\omega = 0.3
    \Gamma$.
    The spacing between constituent meta-atoms in an ASR is $u =
    0.125$.
  }
  \label{fig:magneticMode}
\end{figure}

We find that an incident plane-wave drives all meta-molecules
uniformly and is phase-matched to collective modes in which the
electric and/or magnetic dipoles oscillate in phase.
In the absence of a split-ring asymmetry, only modes involving
oscillating electric dipoles can be driven. These modes strongly emit
perpendicular to the array (into the $\pm\unitvec{e}_z$ directions)
enhancing incident wave reflection.
The magnetic dipoles, however, dominantly radiate into EM field modes
within the ASR plane.
This radiation may become trapped through recurrent scattering
processes in the array, representing modes with suppressed emission
rates and reflectance, and resulting in a transmission resonance.
In order to quantify the effect, we study the radiation properties
(for $\Gamma_{\mathrm{E}} =\Gamma_{\mathrm{M}}$) of the numerically
calculated collective magnetic eigenmode $\colvec{v}_{\mathrm{m}}$ of
the system [figure~\ref{fig:magneticMode}] which maximizes
the overlap
\begin{equation}
  \label{eq:overlap}
  O_{\mathrm{m}}(\colvec{b}_{\mathrm{A}}) \equiv
  \frac{ \left|
      \colvec{v}_{\mathrm{m}}^T \colvec{b}_{\mathrm{A}} \right|^2}
  {\sum_i \left| \colvec{v}_i^T \colvec{b}_{\mathrm{A}} \right|^2}
\end{equation}
with the pure magnetic excitation in which all meta-atom magnetic
dipoles oscillate in phase
\begin{equation}
  \label{eq:bAdef}
  \colvec{b}_{\mathrm{A}}=  \sqrt{\frac{1}{2N}} \left(
    \begin{array}{c}
      +1 \\ -1 \\ \vdots \\+1 \\-1
    \end{array}
  \right)
\end{equation}
The alternating signs between elements of $b_{\mathrm{A}}$ indicate
that the current oscillations of the meta-atoms in an ASR oscillate
antisymmetrically.
We then show that the introduction of an asymmetry $\delta\omega$ in
the resonances allows the excitation of $\colvec{v}_{\mathrm{m}}$ by
the incident field.
This mode closely resembles that responsible for the experimentally
observed transmission resonance
\cite{FedotovEtAlPRL2010,FedotovEtAlPRL2007}.

In an infinite array, each ASR in the magnetic mode would be excited
uniformly, perfectly matching the pure magnetic excitation
$\colvec{b}_{\mathrm{A}}$ in the absence of asymmetry.
This changes in a finite system where boundary effects alter the
distribution of the mode $\colvec{v}_{\mathrm{m}}$.
This is illustrated in figure~\ref{fig:magneticMode} which shows the
numerically calculated uniform magnetic mode $\colvec{v}_{\mathrm{m}}$
in an ensemble of $335$ ASRs with a lattice spacing of $a \simeq 0.28
\lambda$ as in the experiment of Fedotov \emph{et al}
\cite{FedotovEtAlPRL2010}.
To characterize this mode, we examine the electric (symmetric) and
magnetic (antisymmetric) excitations of each ASR.
The state of ASR $\ell$ ($\ell = 1,2,\ldots,N$) is described by the
excitations of its constituent meta-atoms, $b_{2\ell-1}$ and
$b_{2\ell}$.
Therefore, as with a single ASR (\ref{eq:c_pm}), the electric excitation
$c_{\ell,+}$ and the magnetic excitation $c_{\ell,-}$ of ASR
  $\ell$ are given
by the respective symmetric and antisymmetric combinations
\begin{equation}
  \label{eq:c_pm_def}
  c_{\ell,\pm} = \frac{1}{\sqrt{2}} \left(b_{2\ell -1} \pm
    b_{2\ell}\right) \, \textrm{.}
\end{equation}
The magnetic mode excitation consists largely of the ASR magnetic
dipoles oscillating in phase.
In the absence of asymmetry the uniform magnetic mode in a regular
array of ASRs is almost exclusively magnetic in nature
\cite{JenkinsLongPRB}.
Figure~\ref{fig:magneticMode} shows that, on the other hand, an
asymmetry of $\delta\omega = 0.3\Gamma$ provides a small electric
dipole excitation to the magnetic mode, allowing it to be addressed by
the incident field.

\begin{figure}
  \includegraphics{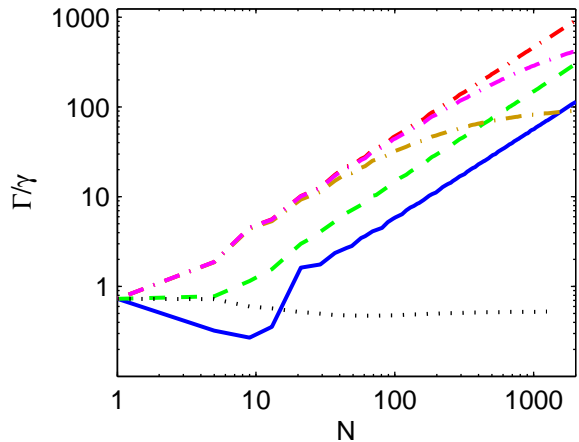}
  \centering
  \caption{Collective resonance narrowing. The resonance
    linewidth $\gamma$ of the collective magnetic mode $\colvec{v}_{\mathrm{m}}$
    in the units of an isolated meta-atom linewidth
    $\Gamma$ as a function of the number
    of meta-molecules $N$; the lattice  spacings $a=1/4\lambda$ (solid
    line), $3/8\lambda$  (dashed line),  $1/2\lambda$ (dot-dashed
    line), and $3/2\lambda$  (dotted line).
    The magenta (intermediate) dot-dashed line corresponds to an asymmetry
    $\delta\omega  = 0.1\Gamma$, while $\delta\omega=0$ for all other
    curves.
    The orange (lower) dot-dashed line incorporates nonradiative loss
    $\Gamma_O=0.01\Gamma$ with all other curves assuming
    $\Gamma_O=0$.}
  \label{fig:linewidth_theory}
\end{figure}

Figure \ref{fig:linewidth_theory} shows the dependence of the resonance
linewidth $\gamma$ of the collective magnetic mode
$\colvec{v}_{\mathrm{m}}$
on the number of meta-molecules $N$ for different
lattice spacings $a$ ($\lambda$ denotes the wavelength of a resonant
incident field).
In the absence of ohmic losses and for sufficiently small
$\delta\omega$, $\gamma\propto 1/N$ for large $N$
when the lattice spacing $a\lesssim \lambda$.
The split ring asymmetry only weakly affects
$\colvec{v}_{\mathrm{m}}$.
For $\delta\omega=0.01 \Gamma$,  the curves representing $\gamma$ are
indistinguishable from those for $\delta\omega=0$.
For the relatively large $\delta\omega=0.1 \Gamma$, however,
$\gamma$ is increased for $N > 200$.
This reduction in quality factor for larger $N$ results from
  the mixing of electric dipoles into the magnetic dipole mode (see
  figure~\ref{fig:magneticMode}), allowing the collective mode to
  emit in the forward and backward, $\pm \unitvec{e}_z$, directions.
The cooperative response and linewidth narrowing sensitively depends
on the lattice spacing $a$.
For larger $a$ (e.g., $a=3/2\lambda$), $\gamma$ becomes insensitive to
$N$, indicating the limit of independent scattering of isolated
meta-molecules and a diminished role of cooperative effects.

The collective behaviour can be understood because the ASR magnetic dipoles emit
  largely into the plane of the metamaterial and the bulk of the
  magnetic mode excitation lies in its interior.
  Any energy radiated from magnetic dipoles would preferentially come
  from the ASRs near the edge of the metamaterial since radiation
  emitted from an ASR on the interior would more likely be
  re-scattered by other ASRs.
  On the other hand, the fraction
  of ASRs at the boundary varies inversely with $N$ leaving an ever
  larger proportion of the magnetic mode $\colvec{v}_{\mathrm{m}}$ in
  the interior of the array for large $N$.
  In the limit of an infinite array ($N,r_c \rightarrow \infty$),
  $\gamma$ would be zero in the  absence of asymmetry $\delta\omega$
  and ohmic losses, for $a\lesssim \lambda$.
  For a sub-wavelength lattice spacing, the only Bragg diffraction
  peak that the array could emit into corresponds to the forward and
  backward scattered fields since all other Bragg peaks would be
  evanescent; but both forward and backward emission from the magnetic
  dipoles are forbidden owing to their orientation.

\begin{figure}
  \includegraphics{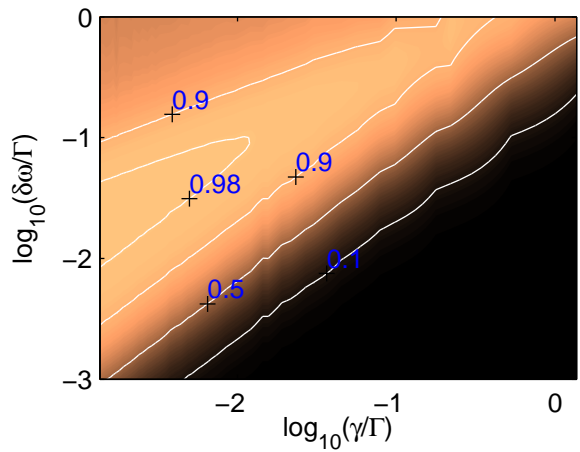}
  \centering
  \caption{The overlap $O_m(\colvec{b}_{\mathrm{f}})$ of
    $\colvec{v}_{\mathrm{m}}$ with the
    state $\colvec{b}_{\mathrm{f}}$ excited by an incident field resonant on the
    mode $\colvec{v}_{\mathrm{m}}$ (for $u=1/2\lambda$, and
    $\Gamma_O=0$). For $\gamma
    \ll \Gamma$ there is a range of asymmetries $\delta\omega$ for
    which the incident field almost exclusively excites the mode
    $\colvec{v}_{\mathrm{m}}$.}
  \label{fig:Overlap}
\end{figure}
An asymmetry $\delta\omega \neq 0$
  generates an effective coupling between the electric and magnetic
  dipoles of individual ASRs in an array [See (\ref{eq:c_pm_evolve})].
  This coupling produces a slight mixing of electric
  dipoles into the phase matched magnetic mode
  $\colvec{v}_{\mathrm{m}}$ of the array as illustrated in
  figure~\ref{fig:magneticMode}.
  We show in figure~\ref{fig:Overlap} that the slight mixing of
  electric dipoles into $\colvec{v}_{\mathrm{m}}$ permits its
  excitation by a uniform resonant driving field propagating
  perpendicular to the plane of the array.
  We represent the steady state excitation of the array induced by
  this field as $\colvec{b}_{\mathrm{f}}$.
  Figure~\ref{fig:Overlap} shows the relative population
  $O_m(\colvec{b}_{\mathrm{f}})$ [See (\ref{eq:overlap})] of the
  phase-matched magnetic mode $\colvec{v}_{\mathrm{m}}$.
 This population
 represents the
  fraction of $\colvec{b}_{\mathrm{f}}$ that resides in the mode
  $\colvec{v}_{\mathrm{m}}$.
We find that for $\gamma\ll\Gamma$ and $\delta\omega \gtrsim \gamma$,
one can induce a state  in which
more than $98\%$ of the energy is in the target mode $\colvec{v}_{\mathrm{m}}$.
For $\delta\omega \ll \gamma$, any excitation that ends up in
$\colvec{v}_{\mathrm{m}}$ is radiated away before it can accumulate; the array
behaves as a collection of radiating electric dipoles.
For larger $\delta\omega$, the population of $\colvec{v}_{\mathrm{m}}$ decreases
since the increased strength of the coupling between
ASR electric and magnetic dipoles begins to excite other modes with nearby
resonance frequencies.
Although the density of modes which may be driven increases linearly
with $N$,  the corresponding reduction of $\gamma$
means that a smaller $\delta\omega$ is needed to excite the target
mode, and there is a range of asymmetries for which $\colvec{v}_{\mathrm{m}}$ is
populated.

\begin{figure}
  \includegraphics{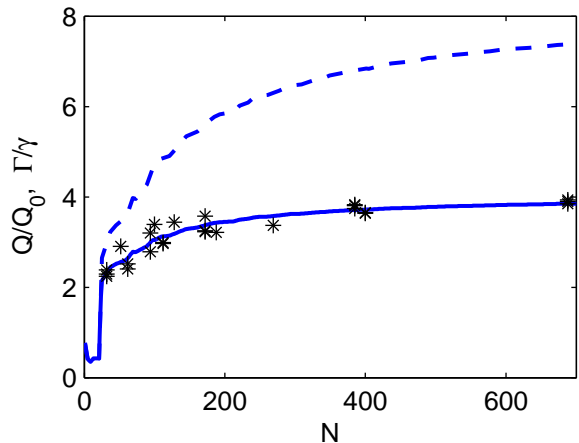}
  \centering
  \caption{Comparison between experimentally measured transmission
    resonance quality factors $Q/Q_0$
    (stars) from Ref.~\cite{FedotovEtAlPRL2010}, where $Q_0\simeq4.5$
    denotes the single meta-atom quality factor
    (for $\lambda\simeq 2.7$cm),
    and numerically calculated
    resonance linewidth $\gamma$ of the collective magnetic mode
    $\colvec{v}_{\mathrm{m}}$ with $\Gamma_{\mathrm{O}}\simeq
    0.14\Gamma$ (solid line) and $\Gamma_{\mathrm{O}}=0$ (dashed
    line).
    Here $\delta\omega\simeq 0.3\Gamma$, and $a\simeq0.28\lambda$. }
  \label{fig:linewidthExp}
\end{figure}
The  narrowing in $\gamma$ combined with the near exclusive excitation
of this mode implies that for larger arrays the radiation from the
sheet is suppressed and hence the transmission enhanced as seen by
Fedotov \textit{et al} \cite{FedotovEtAlPRL2010}.
In figure~\ref{fig:linewidthExp} we compare the experimentally
observed transmission resonance~\cite{FedotovEtAlPRL2010} to our
numerics.
We use the experimental spacing $a\simeq 0.28\lambda$.
The numerical values for the asymmetry $\delta\omega\simeq
  0.3\Gamma$ and the ratio between electric and magnetic spontaneous
  emission rates $\Gamma_{\mathrm{E}}/\Gamma_\mathrm{M} \simeq 1$ were
  estimated from the relative sizes of the ASR meta-atoms
  \cite{FedotovEtAlPRL2007} and their relationship to the resonant 
  wavelength in the experiments by Fedotov \textit{et al}
  \cite{FedotovEtAlPRL2010}.
 The spacing between the meta-atoms $u\simeq 0.125\lambda$ was chosen
 so that the resonance frequencies of the mode $\colvec{v}_{\mathrm{m}}$
 and the collective mode in which all electric dipoles oscillate in
 phase are shifted by less than $\Gamma$ with respect to one and
 other so as to be consistent with experimental observations in
 \cite{FedotovEtAlPRL2010}.
 For a given non-zero $\Gamma_{\mathrm{O}}$, one can obtain the
 collective decay rate of the magnetic mode by adding
 $\Gamma_{\mathrm{O}}$ to the decay rate calculated in the absence of
 ohmic losses.
 We therefore fit the shape of the numerically calculated curve of the
 resonance linewidth as a function of the number of resonators to the
 experimental observations of Fedotov \textit{et al}
 \cite{FedotovEtAlPRL2010} using $\Gamma_{\mathrm{O}}$ 
 as a fitting parameter.
 The ohmic loss rate $\Gamma_{\mathrm{O}} \simeq 0.14
 \Gamma$ produces the expected saturation of quality factor for large
 $N$.
 The vertical shift of the experimental data
   set is determined by the single meta-atom quality factor
 $Q_0 \equiv \omega_0/\Gamma$, and the best value $Q_0\simeq 4.5$ is roughly consistent with full numerical solutions to Maxwell's equations for
 scattering from a single ASR presented by Papasimakis \textit{et
   al.}~\cite{papasimakis2009}. 
The excellent agreement of our simplified model that only includes
dipole radiation contributions from
each meta-atom can be understood by a notably weaker quadrupole than
dipolar radiation field from an ASR~\cite{PapasimakisComm}.

The result also confirms the importance of the uniform magnetic mode
$\colvec{v}_{\mathrm{m}}$ on the observed transmission resonance.
The observed saturation is due to a combination of a fixed
$\delta\omega$, which in larger arrays leads to the population of
  several other modes in addition to $\colvec{v}_{\mathrm{m}}$,
and ohmic losses in the
resonators which set an ultimate limit to the narrowing of $\gamma$.
If ohmic losses were to be reduced, the quality factor of the
resonance would saturate at a correspondingly higher value, as shown in
figure \ref{fig:linewidthExp}.
In the displayed case the resonance linewidth
narrowing is limited by the relatively large asymmetry
$\delta\omega\simeq0.3\Gamma$.
For larger arrays, figure~\ref{fig:linewidth_theory} indicates that
one could further enhance the quality factor by reducing the
asymmetry $\delta\omega$.
So long as $\delta\omega$ sufficiently exceeds the magnetic mode's
  decay rate $\gamma$, a large, and potentially greater,
  fraction of the metamaterial excitation would reside in the magnetic
  mode $\colvec{v}_m$ as shown in figure~\ref{fig:Overlap}.
  The reduced excitation of other collective modes could then further
  enhance the quality of the observed transmission resonance.
A reduction or elimination of ohmic losses would therefore
dramatically increase the resonance's quality factor in larger
arrays.

\section{Conclusion}
\label{sec:conclusion}

In conclusion, we analyzed the recent observations of transmission
spectra in a metamaterial array of ASRs.
We showed that the system can exhibit a strong cooperative response in the
case of sufficiently closely-spaced resonators.
Moreover, we demonstrated how an asymmetry in the split rings leads to
excitation of collective magnetic modes by a field which does not
couple directly to ASR magnetic moments.
The excitation of this uniform phase-coherent mode results in
cooperative response exhibiting a dramatic resonance linewidth
narrowing, explaining the experimental
findings~\cite{FedotovEtAlPRL2010}.

\ack
We gratefully acknowledge N.\ Papasimakis, V.\ Fedotov, and N.\
Zheludev for discussions and providing the measurement data for the
quality factor. This work was financially supported by the EPSRC and
the Leverhulme Trust.

\section*{References}


\end{document}